\pdfoutput=1
\PassOptionsToPackage{unicode}{hyperref}
\PassOptionsToPackage{hyphens}{url}
\documentclass[
]{statsoc}
\author{}
\date{\vspace{-2.5em}}
\usepackage{footnote}
\usepackage{amsmath,amssymb}
\usepackage{float}
\usepackage{lmodern}
\usepackage{iftex}
\ifPDFTeX
  \usepackage[T1]{fontenc}
  \usepackage[utf8]{inputenc}
  \usepackage{textcomp} 
\else 
  \usepackage{unicode-math}
  \defaultfontfeatures{Scale=MatchLowercase}
  \defaultfontfeatures[\rmfamily]{Ligatures=TeX,Scale=1}
\fi
\IfFileExists{upquote.sty}{\usepackage{upquote}}{}
\IfFileExists{microtype.sty}{
  \usepackage[]{microtype}
  \UseMicrotypeSet[protrusion]{basicmath} 
}{}
\makeatletter
\@ifundefined{KOMAClassName}{
  \IfFileExists{parskip.sty}{%
    \usepackage{parskip}
  }{
    \setlength{\parindent}{0pt}
    \setlength{\parskip}{6pt plus 2pt minus 1pt}}
}{
  \KOMAoptions{parskip=half}}
\makeatother
\usepackage{xcolor}
\IfFileExists{xurl.sty}{\usepackage{xurl}}{} 
\IfFileExists{bookmark.sty}{\usepackage{bookmark}}{\usepackage{hyperref}}
\hypersetup{
  hidelinks,
  pdfcreator={LaTeX via pandoc}}
\usepackage{graphicx}
\makeatletter
\def\maxwidth{\ifdim\Gin@nat@width>\linewidth\linewidth\else\Gin@nat@width\fi}
\def\maxheight{\ifdim\Gin@nat@height>\textheight\textheight\else\Gin@nat@height\fi}
\makeatother
\setkeys{Gin}{width=\maxwidth,height=\maxheight,keepaspectratio}
\makeatletter
\def\fps@figure{htbp}
\makeatother
\title{Statistical Modeling of Networked Evolutionary Public Goods Games}

\author[Ando et al.]{Hiroyasu Ando}
\coaddress{Hiroyasu Ando, Department of Biostatistics, University of California, Los Angeles, California, 90095, USA. Email: hiro1999@ucla.edu}
\address{Department of Biostatistics, University of California, Los Angeles, California, 90095, USA.}

\author[Ando et al.]{Akihiro Nishi}
\address{Department of Epidemiology, University of California, Los Angeles, California, 90095, USA}

\author[Ando et al.]{Mark S. Handcock}
\address{Department of Statistics \& Data Science, University of California, Los Angeles, California, 90095, USA}

\usepackage{amsmath}\usepackage{amsfonts}\usepackage{amssymb}\usepackage{mathrsfs}\usepackage{graphicx}\usepackage{float}\usepackage{url}\usepackage{epstopdf}\usepackage{booktabs}\usepackage{longtable}\usepackage{appendix}\usepackage[a4paper]{geometry}\usepackage{graphicx}\usepackage[textwidth=8em,textsize=small,colorinlistoftodos]{todonotes}\usepackage{amsmath}\usepackage{natbib}\usepackage{hyperref}\usepackage[ruled,vlined]{algorithm2e}\usepackage{xcolor}\usepackage{colortbl}\usepackage{bigdelim}\usepackage{breqn}
\usepackage{booktabs}
\usepackage{longtable}
\usepackage{array}
\usepackage{multirow}
\usepackage{wrapfig}
\usepackage{float}
\usepackage{colortbl}
\usepackage{pdflscape}
\usepackage{tabu}
\usepackage{threeparttable}
\usepackage{threeparttablex}
\usepackage[normalem]{ulem}
\usepackage{makecell}
\usepackage{xcolor}
\usepackage{threeparttable} 
\ifLuaTeX
  \usepackage{selnolig}  
\fi
\usepackage{lineno}

\begin{document}

\newtheorem{assumption}{Assumption}
\newtheorem{definition}{Definition}
\newcommand{\R}{\mathbb{R}}
\newcommand{\N}{\mathbb{N}}
\newcommand{\E}{\mathbb{E}}
\newcommand{\V}{\mathbb{V}}
\newcommand{\bfR}{\mathbf{R}}
\newcommand{\bfX}{\mathbf{X}}
\newcommand{\bfW}{\mathbf{W}}
\newcommand{\bfD}{\mathbf{D}}
\newcommand{\INT}{\int_{-\infty}^{+\infty}}
\newcommand{\p}{\partial}
\newcommand{\ra}{\Rightarrow}
\newcommand{\dH}{d\mathscr{H}}
\newcommand{\ch}{\text{cosh}}
\newcommand{\sh}{\text{sinh}}
\newcommand{\ex}{\mathbb{E}\left[X\right]}
\newcommand{\ey}{\mathbb{E}\left[Y\right]}
\newcommand{\indep}{\perp \!\!\! \perp }

\setcounter{secnumdepth}{4}

\begin{abstract}
Repeated small dynamic networks are integral to studies in
evolutionary game theory, where networked public goods games offer novel insights into human behaviors. Building on these findings, it is necessary to develop a statistical model that effectively captures dependencies across multiple small dynamic networks. While Separable Temporal Exponential-family Random Graph Models (STERGMs) have demonstrated success in modeling a large single dynamic network, their application to multiple small dynamic networks with less than 10 actors, remains unexplored.  In this study, we extend the STERGM framework to accommodate multiple small dynamic networks, offering an approach to analyzing such systems.  Taking advantage of the small network sizes, our proposed approach improves accuracy in statistical inference through direct computation, unlike conventional approaches that rely on Markov Chain Monte Carlo methods.  We demonstrate the validity of this framework through the analysis of a networked public goods experiment into individual decision-making about cooperation and defection. The resulting statistical inference uncovers novel insights into the dynamics of social dilemmas, showcasing the effectiveness and robustness of this modeling and approach.

\end{abstract}
\keywords{Evolutionary Game Theory; Experimental Game Theory; Longitudinal networks; Public Goods Game; Social Networks; Separable Temporal Exponential-family Random Graph Models.}

\section{Introduction}\label{sec:intro}

Networks are widely used to represent relational information, enabling a deeper understanding of structures and dependencies within social relationships. In addition, there is increasing interest in using dynamic networks to represent relational information evolving over time in various fields.  In evolutionary game theory, networked public goods games have attracted much attention for their ability to shed light on human behaviors. For instance, using networked public goods games, \citet{ref18} demonstrated that wealth visibility leads to lower cooperation rates and inter-connectedness.

In traditional public goods games, individuals repeatedly encounter a choice between cooperation (benefiting the group) and defection (benefiting themselves) among 4 to 6 participants \citep{ref5, ref17, ref19}. Recent adaptations of public goods games incorporate network structures to imitate realistic social interactions \citep{ref18, ref22, ref3}.  In networked settings, public goods games generate multiple dynamic small networks where the temporal relational information reflects repeated interactions.

As highlighted above, public goods games have a long history and have been extensively studied. However, despite advances in the field, many studies continue to focus on nodal-level analysis, which may overlook the critical dependencies within networks that are essential for understanding broader dynamics. Building on these findings, there is a clear need to develop a modeling framework that effectively captures dependencies across multiple small dynamic networks. Such an approach could provide deeper insights into the complexities of networked human behaviors and enhance our understanding of relational dynamics over time.

Separable Temporal Exponential-family Random Graph Models (STERGMs) are an established method for representing dependencies in temporal relational information. These models are widely used for modeling single large dynamic networks, typically with 20 nodes or more \citep{ref13, ref16, ref15}. However, STERGM applications in smaller multiple networks—though common in real-world contexts—remain relatively unexplored. Such repeated small dynamic networks are integral to studies in evolutionary game theory, where networked public goods games offer novel insights into networked human behaviors in social dilemmas. 

In this paper, we develop families of STERGMs to represent the complex relational structures observed in networked public goods games on evolutionary game theory. In Section 2, we provide an overview of the public goods game employed in our study. Section 3 introduces and extends STERGMs to accommodate the context of multiple small dynamic networks. Finally, in Section 5, we apply this modeling approach to networked public goods games and perform statistical inference to uncover key insights.

\section{Networked Public Goods Games}\label{sec5}

In this section we describe the structure of a typical experiment on a networked public goods game. This description exposes the canonical characteristics of such experiments and games.

\subsection{Game structure\label{subsec5.1}}

Each game consisted of 6 participants who were mutually identified through randomly assigned and unique labels 1-6. 
Participants were each given an initial wealth of 500 units and randomly assigned to a game with 5 other participants. In each game, five network ties were formed randomly in an Erdős-Rényi design \citep{ref4}. Participants who were tied were referred to as neighbors. Participants took part in a 7-round process where their cumulative wealth was converted into real money at the end of the game at a rate of 1000 units per dollar. Notably, participants were unaware of the total number of rounds they would play.

\begin{figure} [H]
\centering
\includegraphics[width=0.8\textwidth]{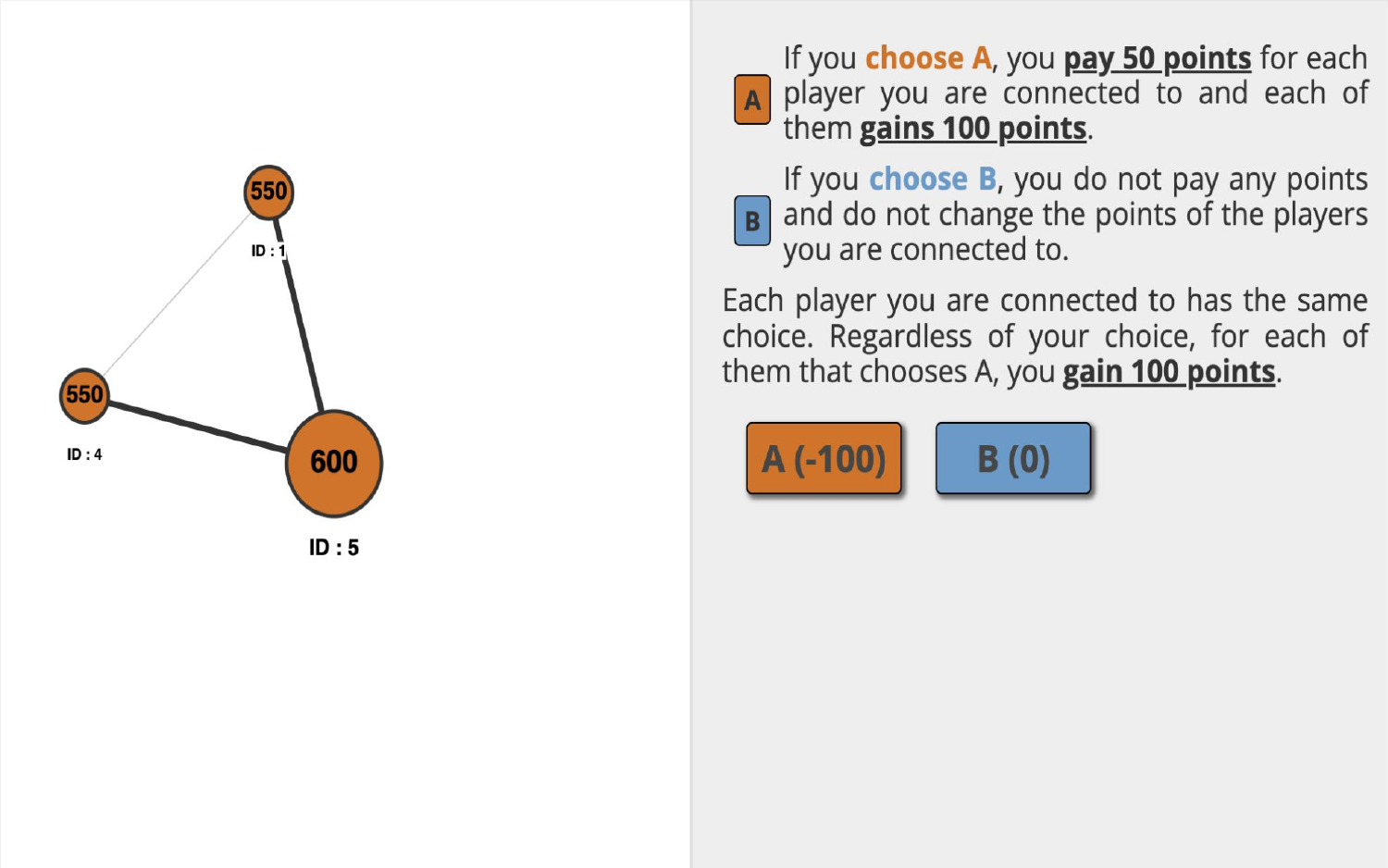} 
\caption{Decision Phase in the Networked Public Goods Games\label{fig1}}
\end{figure}

\begin{figure} [H]
\centering
\includegraphics[width=0.8\textwidth]{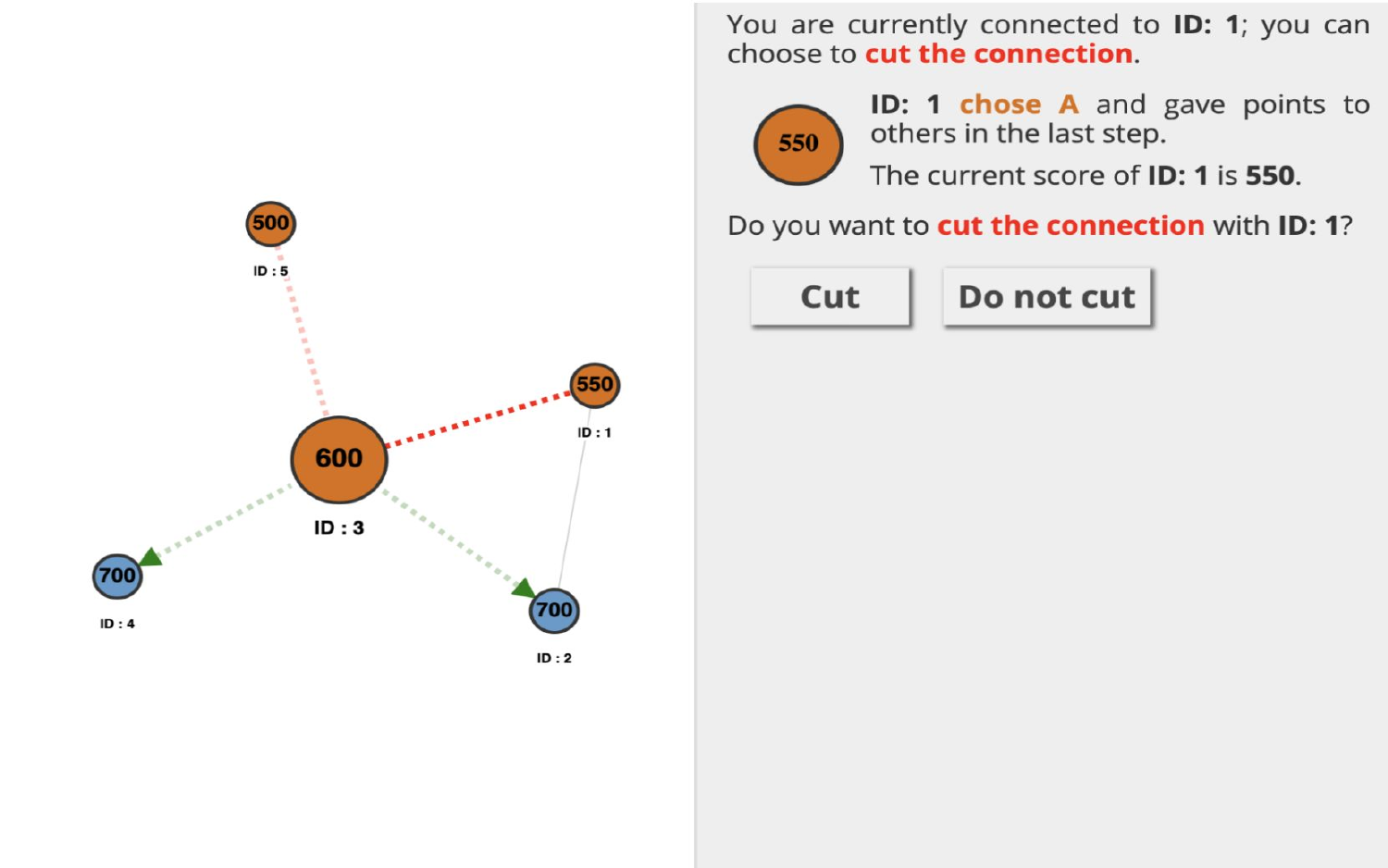}
\caption{Dissolving ties in the Networked Public Goods Games\label{fig2}}
\centering
\includegraphics[width=0.8\textwidth]{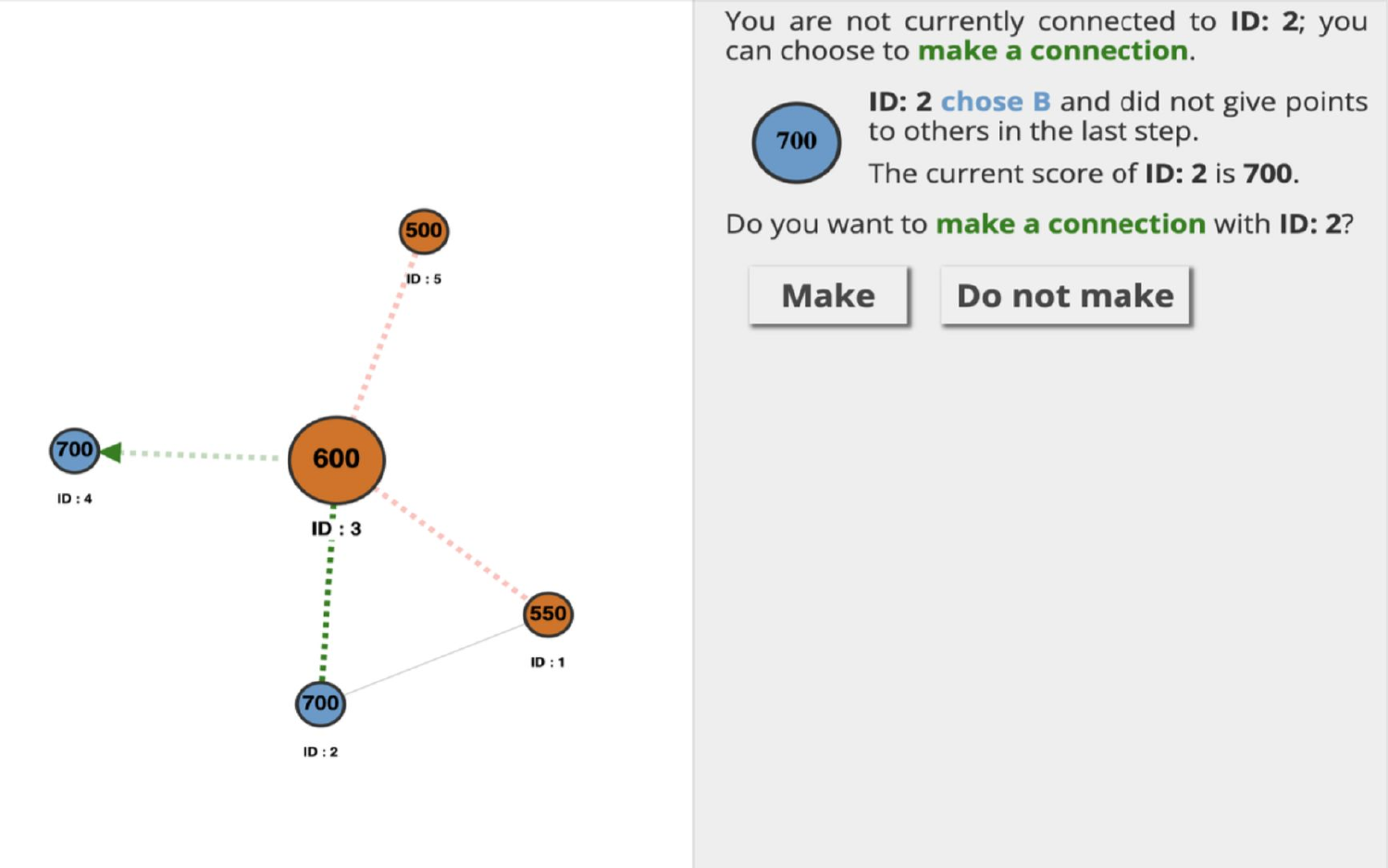}
\caption{Forming ties in the Networked Public Goods Games\label{fig3}}
\centering
\end{figure}

During each round, participants faced a choice between cooperation and defection (Fig. \ref{fig1}).  In cooperation, participants reduce their own wealth by 50 units per neighbor to increase the wealth of each neighbor by 100 units.  In defection, participants retain their wealth while providing no benefit to their neighbors. Each participant was allowed a single decision that applied uniformly to all of their neighbors. After making their choices, participants received information about their neighbors’ decisions and were allowed to change their ties. Specifically, 5 tied and untied randomly selected of participants were given opportunities to change their ties. If a tie already existed between two participants, it could be dissolved if either participant chose to dissolve it (Fig. \ref{fig2}).  If no tie existed, a new tie could be formed if both participants chose to form it (Fig. \ref{fig3}).

At the beginning of the experiment, participants played with AI counterparts for two practice rounds.  Following this, the main game commenced.

Several key features shaped participants' decision-making and influenced network dynamics throughout the game. At each point each participant was able to recognize their neighbors and recall the history of their interactions, decisions and wealth.  Participants could see their own cumulative wealth points as well as those of their neighbors, providing wealth visibility.  Behavioral transparency allowed neighbors to observe the most recent decisions of each participant, whether cooperation or defection.  Additionally, participants had access to the structure of ties between their neighbors, fostering ego-alter network awareness.  Note that during the change phase, participants were unable to perceive ongoing changes in ties made by others.

These features allowed participants to strategically form or dissolve ties based on their neighbors’ identification numbers, wealth points, recent decisions, and the observed structure of ties among neighbors.  For instance, participants could form new ties with cooperative individuals or dissolve ties with defectors, potentially leading to the formation and dissolution of dynamic social networks influenced by economic dependencies and strategic interactions.

The experiments were conducted using Breadboard \citep{ref25}, a software platform designed for online social experiments. Participants were recruited from around the world using Prolific, a platform to recruit participants online, in November 2024.

\subsection{Experiments}

We conducted 20 experiments based on the networked public goods game for
a total of 120 participants across all networks.

\begin{figure} [H]
\centering
\includegraphics[width=0.8\textwidth]{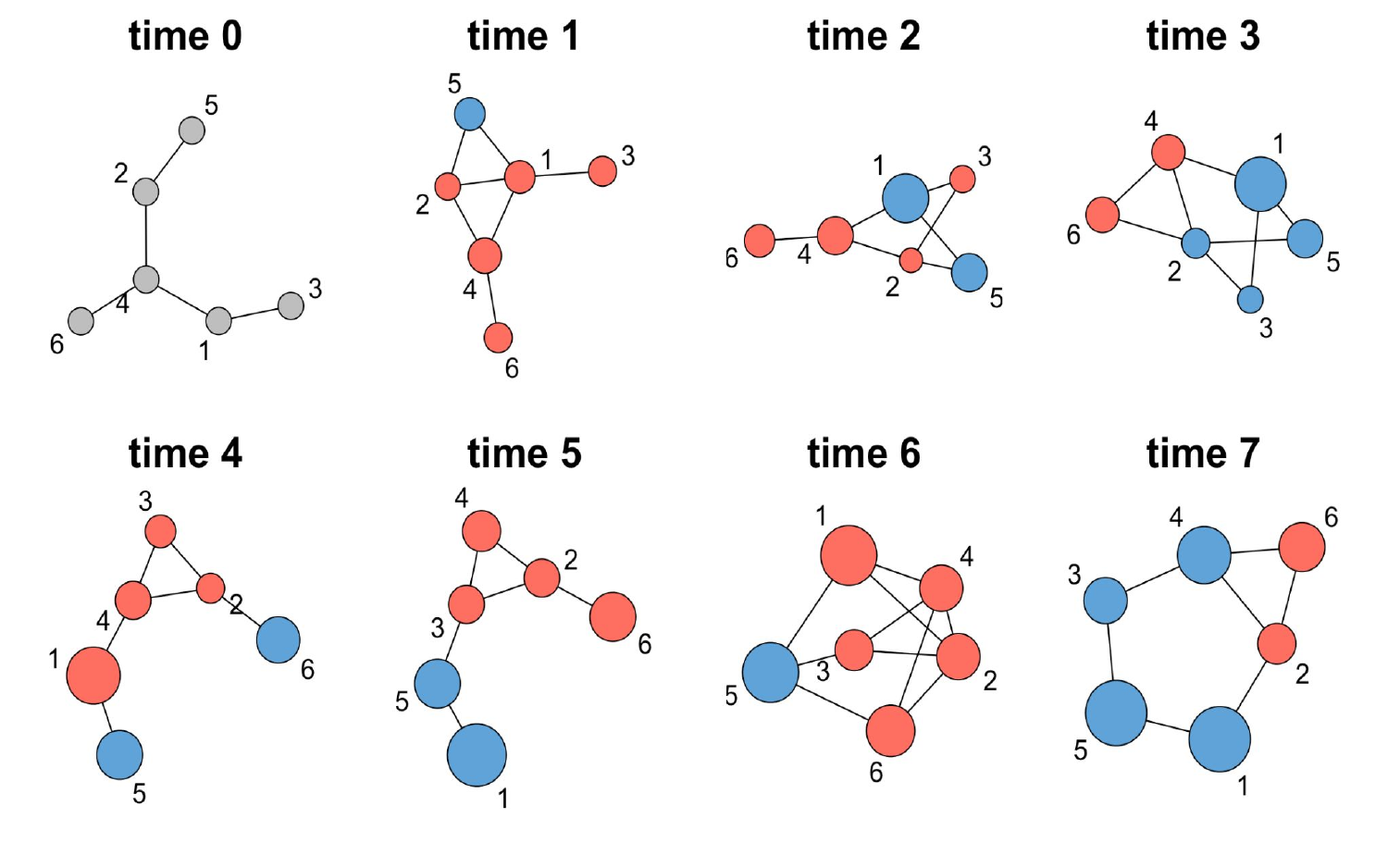}
\caption{Network dynamics in the networked public goods game\label{fig4}}
\end{figure}

Fig. \ref{fig4} shows The networks formed during one of the public goods games. Each node is a participant and the ties indicate connectiveness.  The node size indicates wealth.  The bigger the nodes were richer and vice versa. The numbers 1-6 denote the identification number of each participant.  The node color represents the most recent decisions of each participant, whether cooperation or defection (red: cooperation, blue; defection, grey: no history).  In each time step, the decision phase (cooperation or defection ) and the subsequent change phase are one set.  For example, in time 1, the participant `1' cooperated with its neighbors and as a result, connected with participants `2' and `5'.

\section{Separable Temporal Exponential-family Random Graph Models}\label{sec3}

In this section we describe a statistical model to represent the relationships within the game.  We start with a model for a single network and then expand it to the larger set of $G$ networks.

Consider a longitudinal network of relations between $n$ actors, labeled $\{1, \ldots, n\}$ at time points $t=1, \ldots, T$.  Let $Y^{t}_{ij}$ be a random variable representing a measure of the relation between actors $i$ and $j$ at time $t$, so that the matrix ${\mathbf Y}^{t}=[Y_{ij}]_{n{\times}n}$ can be thought of as a random graph over the set of actors. Let $\mathbf{y}^{t} \in \mathcal{Y}$ be a realization of $\mathbf{Y}^{t}$.

Separable Temporal Exponential-family Random Graph Models were introduced by \citet{ref13} as a subset of temporal exponential-family random graph models for
better interpretability and model specification. The main concept is to ``separate'' the dynamic network into distinct formation and persistence processes.

Consider the network transition from time $t-1$ to time $t$, defining the network $\mathbf{Y}^{t-1}$ at time $t-1$, the network $\mathbf{Y}^{t}$ at time $t$, the formation network $\mathbf{Y}^+$; the initial network $\mathbf{Y}^{t-1}$ with the addition of ties at time $t$, and the persistence network $\mathbf{Y}^-$; the initial network $\mathbf{Y}^{t-1}$ with the removal of ties at time $t$. Via a set operation, the realized formation and persistence networks are derived as:

\begin{equation}
\begin{aligned}
  &\mathbf{y}^+ = \mathbf{y}^{t-1} \cup \mathbf{y}^t  \\
  &\mathbf{y}^- = \mathbf{y}^{t-1} \cap \mathbf{y}^t
\end{aligned}
\end{equation}
In this operation, $\mathbf{y}^+ = \mathbf{y}^{t-1} \cup \mathbf{y}^t$ represents the set of ties that appear in either the network at time $t-1$ or the network at time $t$. Conversely, $\mathbf{y}^- = \mathbf{y}^{t-1} \cap \mathbf{y}^t$ represents the set of ties that exist in both the network at time $t-1$ and the network at time $t$. A key goal of STERGMs is to reconstruct $\mathbf{y}^t$ from $\mathbf{y}^{t-1}$, $\mathbf{y}^+$, and $\mathbf{y}^-$, or to separate $\mathbf{y}^t$ into $\mathbf{y}^+$ and $\mathbf{y}^-$, given $\mathbf{y}^{t-1}$. This reconstruction is achieved with the following set operation:
\begin{equation}
\mathbf{y}^t = \mathbf{y}^+ \backslash (\mathbf{y}^{t-1} \backslash \mathbf{y}^-) = \mathbf{y}^- \cup (\mathbf{y}^+ \backslash \mathbf{y}^{t-1}), 
\end{equation}
where, $\mathbf{y}^+ \backslash \mathbf{y}^{t-1}$ contains ties $\{i, j\}$  that are present in $\mathbf{y}^+$ but not in $\mathbf{y}^{t-1}$. Thus, $\mathbf{y}^t$ can be expressed as the union of $\mathbf{y}^-$ and $\mathbf{y}^+ \backslash \mathbf{y}^{t-1}$. This approach allows us to separate the processes of ties into the formation and the persistence as the network evolves over time. As a result, if $\mathbf{Y}^+$ is independent of $\mathbf{Y}^-$ conditional on $\mathbf{Y}^{t-1}$, the transition probability from time $t-1$ to time $t$ is separable as follows:
\begin{align}
&P(\mathbf{Y}^t = \mathbf{y}^t | \mathbf{Y}^{t-1} = \mathbf{y}^{t-1}; \boldsymbol{\theta}) = P(\mathbf{Y}^+ = \mathbf{y}^+, \mathbf{Y}^- = \mathbf{y}^- | \mathbf{Y}^{t-1} = \mathbf{y}^{t-1}; \boldsymbol{\theta}^+, \boldsymbol{\theta}^-) \nonumber \\
&=P(\mathbf{Y}^+ = \mathbf{y}^+ | \mathbf{Y}^{t-1} = \mathbf{y}^{t-1}; \boldsymbol{\theta}^+) \times P(\mathbf{Y}^- = \mathbf{y}^- | \mathbf{Y}^{t-1} = \mathbf{y}^{t-1}; \boldsymbol{\theta}^-).
\end{align}

Specifically, we respectively model the formation and the persistence models. Given $\boldsymbol{y}^{t-1} \in \mathcal{Y}$, the realizations of $\boldsymbol{Y}^+$ can be expressed as $\boldsymbol{y}^+ \in \mathcal{Y}^+(\boldsymbol{y}^{t-1}) \subseteq\{\forall \boldsymbol{y} : \boldsymbol{y} \supseteq \boldsymbol{y}^{t-1}\}$ and the realizations of $\boldsymbol{Y}^-$ is expressed as $\boldsymbol{y}^- \in \mathcal{Y}^-(\boldsymbol{y}^{t-1}) \subseteq\{\forall \boldsymbol{y} : \boldsymbol{y} \subseteq \boldsymbol{y}^{t-1}\}$. With a $d$-vector $g^+(\boldsymbol{y}^{+}, \boldsymbol{y}^{t-1})$ of sufficient statistics for the formation network $\boldsymbol{y}^+$ from $\boldsymbol{y}^{t-1}$ and parameter $\boldsymbol{\theta}^+ \in \mathbb{R}^d$ and a $d$-vector $g^-(\boldsymbol{y}^{-}, \boldsymbol{y}^{t-1})$ of sufficient statistics for the persistence network $\boldsymbol{y}^-$ from $\boldsymbol{y}^{t-1}$ and parameter $\boldsymbol{\theta}^- \in \mathbb{R}^d$, the formation and persistence models are elaborated as:
\begin{align}
P(\boldsymbol{Y}^+ = \boldsymbol{y}^+ | \boldsymbol{Y}^{t-1} = \boldsymbol{y}^{t-1}; \boldsymbol{\theta}^+) = \frac{\exp(\boldsymbol{\theta}^{+} {\cdot}g^+(\boldsymbol{y}^{+}, \boldsymbol{y}^{t-1}))}{c^+(\boldsymbol{\theta}^+, \boldsymbol{y}^{t-1})} \quad \boldsymbol{y}^+ \in \mathcal{Y}^+(\boldsymbol{y}^{t-1}), \\
P(\boldsymbol{Y}^- = \boldsymbol{y}^- | \boldsymbol{Y}^{t-1} = \boldsymbol{y}^{t-1}; \boldsymbol{\theta}^-) = \frac{\exp(\boldsymbol{\theta}^{-} {\cdot}g^-(\boldsymbol{y}^{-}, \boldsymbol{y}^{t-1}))}{c^-(\boldsymbol{\theta}^-, \boldsymbol{y}^{t-1})} \quad \boldsymbol{y}^- \in \mathcal{Y}^-(\boldsymbol{y}^{t-1}),
\end{align}
where 
\begin{align}
c^+(\boldsymbol{\theta}^+, \boldsymbol{y}^{t-1}) &= \sum_{\boldsymbol{x}^+ \in \mathcal{Y}^+(\boldsymbol{y}^{t-1})} \exp\{\boldsymbol{\theta}^{+} {\cdot}g^+(\boldsymbol{x}^{+}, \boldsymbol{y}^{t-1})\}, \\
c^-(\boldsymbol{\theta}^-, \boldsymbol{y}^{t-1}) &= \sum_{\boldsymbol{x}^- \in \mathcal{Y}^-(\boldsymbol{y}^{t-1})} \exp\{\boldsymbol{\theta}^{-} {\cdot}g^-(\boldsymbol{x}^{-}, \boldsymbol{y}^{t-1})\},
\end{align}
are the normalizing constants. In this framework, the sufficient statistics for the formation and persistence networks can vary, allowing for a more flexible model specification \citep{ref13}. In practice, this property is considered to be useful \citep{ref11, ref13}. For instance, in an extreme case, the formation network model might include statistics that capture homophily ties, while the persistence network does not. Although STERGMs sacrifice the ability to model interactions between the formation and persistence networks intra-time step, it offers significant improvements in model specification and interpretability.

We extend STERGMs to model the multiple small dynamic networks that are the result of the evolutionary games.  Suppose we have $G$ independent small dynamic networks from the same experimental setting, each with $T$ time points and $n$ nodes of interest. The nodes in each of the $G$ dynamic networks are distinct. Let $\boldsymbol{Y}^{t,g}$ be an undirected random graph at time $t$ in the $g$-th dynamic network, whose realization is $\boldsymbol{y}^{t,g} \in \mathcal{Y}$, the set of possible networks of interest on $n$.  With a $d$-vector $g(\boldsymbol{y}^{t,g}, \boldsymbol{y}^{t-1,g})$ of sufficient statistics for the network transition from $\boldsymbol{y}^{t-1,g}$ to $\boldsymbol{y}^{t,g}$ and parameter $\boldsymbol{\theta} \in \mathbb{R}^d$, the transition probability from time $t-1$ to time $t$ in the $g$-th network is defined as:

\begin{equation}
P(\boldsymbol{Y}^{t,g}=\boldsymbol{y}^{t,g}|\boldsymbol{Y}^{t-1,g}=\boldsymbol{y}^{t-1,g} ; \boldsymbol{\theta}) = \frac{\exp\{\boldsymbol{\theta}{\cdot}g(\boldsymbol{y}^{t,g}, \boldsymbol{y}^{t-1,g})\}}{c(\boldsymbol{\theta}, \boldsymbol{y}^{t-1,g})} \quad \boldsymbol{y}^{t,g}, \boldsymbol{y}^{t-1,g} \in \mathcal{Y},
\end{equation}
where
\begin{equation}
c(\boldsymbol{\theta}, \boldsymbol{y}^{t-1,g}) = \sum_{\boldsymbol{x}^{t,g} \in \mathcal{Y}} \exp\{\boldsymbol{\theta}g(\boldsymbol{x}^{t,g}, \boldsymbol{y}^{t-1,g})\}
\end{equation}
is the normalizing constant. As a result, assuming homogeneity of parameters over time and networks, the likelihood of a STERGM with $G$ independent networks and $T$ time points can be represented as: 
\begin{equation}
\prod_{g=1}^{G}\prod_{t=2}^{T}P(\boldsymbol{Y}^{t,g}=\boldsymbol{y}^{t,g}|\boldsymbol{Y}^{t-1,g}=\boldsymbol{y}^{t-1,g} ; \boldsymbol{\theta})\label{eq:lik}.
\end{equation}
This framework is a natural extension of STERGMs, retaining the same interpretability.  Statistical inferences can be conducted using Markov Chain Monte Carlo (MCMC) methods.  However, leveraging the small network size, especially $n \leq 7$, inherent to multiple small networks, it becomes feasible to numerically calculate the likelihood function directly and estimate parameters with the direct numerical optimization.  This approach has significant advantages, including producing more reliable parameter estimates, standard errors, and the likelihood ratios, allowing for robust model comparisons through the deviance test.  In contrast, the MCMC approaches often encounter challenges such as poorly mixed chains, the presence of MCMC errors, and uncertainties in approximating likelihood ratios \citep{ref10}.

\subsection{Statistical inference \label{subsec5.2}}

We aimed to investigate the dynamic structural patterns stemming from human behaviors in social dilemmas using networked public goods games. To achieve this, we focused on modeling the formation and persistence of networks by employing the proposed model with the STERGM parameterization. This approach assumes homogeneity of parameters across time and games. Following \citet{ref13}, we incorporated both exogenous and endogenous structural statistics. Note that identical statistics were used for both the formation and persistence models, as the network dynamics were presumed to be governed by the same structural patterns (albeit with different parameters). The dynamic networks were undirected due to the symmetric interactions inherent in the public goods game.

First, we included terms for the number of cooperation and defection homophily ties. For instance, at each step, a cooperation homophily tie was defined after the change phase if both a participant and its neighbor had previously chosen cooperation. These statistics capture the dependencies arising from participants’ strategic decisions. Second, we incorporated terms for the sum of absolute wealth differences between tied participants. This statistic reflects economic dependencies, capturing how wealth disparities influence decisions to form or dissolve ties. Finally, we included terms for the number of edges and triangles to account for broader structural patterns in the dynamics networks. Unlike studies on directed networks \citep{ref13, ref24}, we omitted terms for aggregate transitive and cyclical ties, as our data involved undirected networks without hierarchical interactions.

We computed the maximum likelihood estimate (MLE) by direct optimization
using the Broyden-Fletcher-Goldfarb-Shanno (BFGS) algorithm \citep{ref2}. We can estimate the MLE without Markov Chain Monte Carlo (MCMC) \citep{ref10} or other approximations, such as pseudo-likelihood \citep{str90}, as we are leveraging the small network size of our multiple small dynamic networks. Model comparisons were conducted using likelihood ratio tests, enabling evaluation of model fit and selection of the most suitable model for the networked public goods game.

The validity of the MLE for this setting is based on two arguments. The first are studies for the MLE for ERGM in small network size settings \citep{yonetal21}. They find that the MLE is a good estimator even for small network sizes. The second evidence comes from asymptotics: as the number of experiments, $G$, increases the MLE satisfies a central limit theorem. Specifically, under mild regularity conditions, the MLE exists with probability approaching one, is unique when it exists and is asymptotically Gaussian with mean the true value of the parameter and covariance equal to the inverse Fisher information matrix (corresponding to the likelihood in equation $(\ref{eq:lik})$ \citep{Barndorff1978, geyer2013expfam}.
In our situation, we compute the information matrix numerically from the Hessian returned as a by-product of the optimization.
Results by \citet{bogdan2022mle} suggest that the asymptotics is relevant if $G$ is of the same size as the number of nodes, $n$. In our situation, $G=20$ and $n=7$.

Table \ref{tab1} presents the model estimates, and we provide brief interpretations for significant parameters below. For the formation model, the edge parameter was estimated at -1.358 (SE = 0.195), this indicates that the formation of ties is less likely than in homogeneous networks, controlling for covariates and structural dependencies. The triangle parameter was estimated at -0.260 (SE = 0.096), suggesting that ties completing triangles are less likely to form compared to ties that do not form such structures, controlling for covariates and structural dependencies. The cooperation homophily parameter was estimated at 1.069 (SE = 0.183), showing that ties between cooperative participants are more likely to form than heterophilous ties, controlling for other covariates and structural dependencies. The wealth difference parameter was estimated at -1.084 (SE = 0.542), this implies that ties are less likely to form between participants with larger wealth disparities compared to those with smaller differences, controlling for other covariates and structural dependencies.

For the persistence model, the edge parameter was estimated at 1.682 (SE = 0.210), indicating that existing ties are more likely to persist compared to homogeneous networks, controlling for covariates and structural dependencies. The cooperation homophily parameter was estimated at 1.677 (SE = 0.276), showing that ties between cooperative participants are more likely to persist than heterophilous ties, controlling for other covariates and structural dependencies.  The wealth difference parameter was estimated at -1.099 (SE = 0.469), this implies that ties are less likely to persist between participants with larger wealth disparities compared to those with smaller differences, controlling for other covariates and structural dependencies.

\begin{table}
\caption{MLE parameter estimates for the public goods game networks\label{tab1}}
\begin{tabular}{lll}
\toprule
Parameter & Formation    & Persistence \\
          & est. (s.e.)  & est. (s.e.) \\
\midrule
Edge                       & \text{-1.358} (0.195)***  & \text{ 1.682 (0.210)}*** \\
Triangle                   & \text{-0.260} (0.096)**   & \text{-0.023 (0.133)}   \\
Homophily (cooperation)    & \text{ 1.069} (0.183)***  & \text{ 1.677 (0.276)}*** \\
Homophily (defection)      & \text{ 0.438} (0.255)     & \text{ 0.106 (0.225)}    \\
Absolute wealth difference & \text{-1.084} (0.542)*    & \text{-1.099 (0.469)}*   \\  
\bottomrule
\addlinespace[1ex]
\multicolumn{3}{l}{Significance levels: 0.05*, 0.01**,  0.001***}
\end{tabular}
\end{table}

\begin{table}
\caption{Deviances for the public goods game networks\label{tab2}}
\begin{tabular}{llll}
\toprule
Model & Residual deviance & Deviance    &  p value \\
      &                   & dev. (d.f.) &         \\
\midrule
Null            & 2911.22          & ---               & ---     \\
Only-Covariates & 1742.47          & 1168.74 (8)      & 0.000   \\
Full            & 1734.79          & 7.68~~~~{\kern1.5pt}~(2) & 0.021   \\
\bottomrule
\end{tabular}
\end{table}

Our analysis revealed key insights into the dynamics of public goods games.  There is a negative transitive effect in the formation of ties. Participants might tend to avoid forming ties that complete triangles. This pattern suggests strategic considerations may directly or indirectly drive avoidance of transitive clustering, particularly in such competitive environments. There is a tendency for participants to form and maintain ties with others of similar wealth levels. This highlights the economic disparities in both the formation and persistence of ties. Consistent with expectations, a strong preference for forming and persisting ties between cooperative participants was observed. This might reflect the influence of underlying social norms, emphasizing the importance of cooperative behavior in fostering stable and cohesive networks within the social dilemma.

Table \ref{tab2} provides the deviance test for the null model, only-covariates model, and discussed (full) model. We see that the discussed (full) model significantly improved the fit, compared to the null model and only-covariates model. This reveals that the structural term, the number of triangles, plays an important role to explain the network dynamics. 

\begin{figure} [H]
\centering
\includegraphics[width=\textwidth]{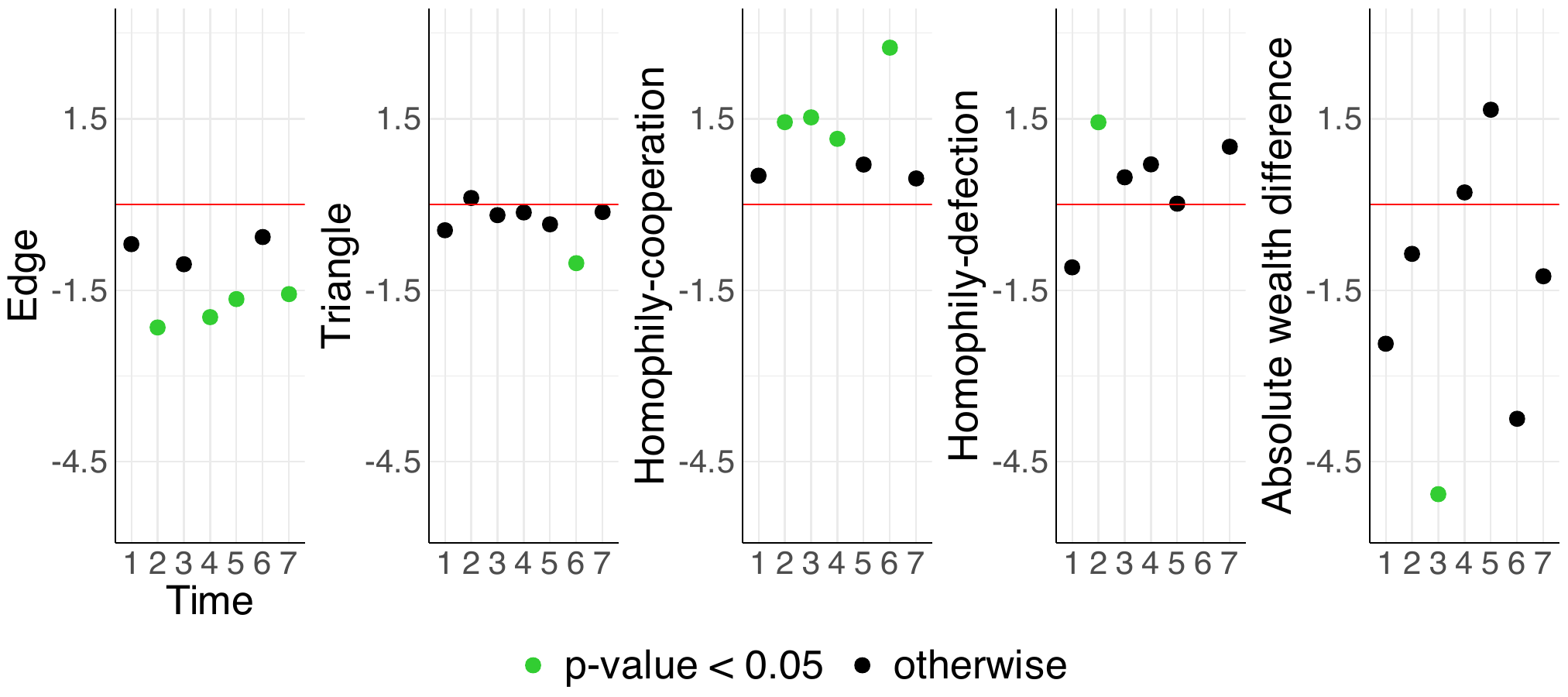}
\caption{MLE parameter estimates for the formation model over time\label{fig5}}

\vspace{2em}

\centering
\includegraphics[width=\textwidth]{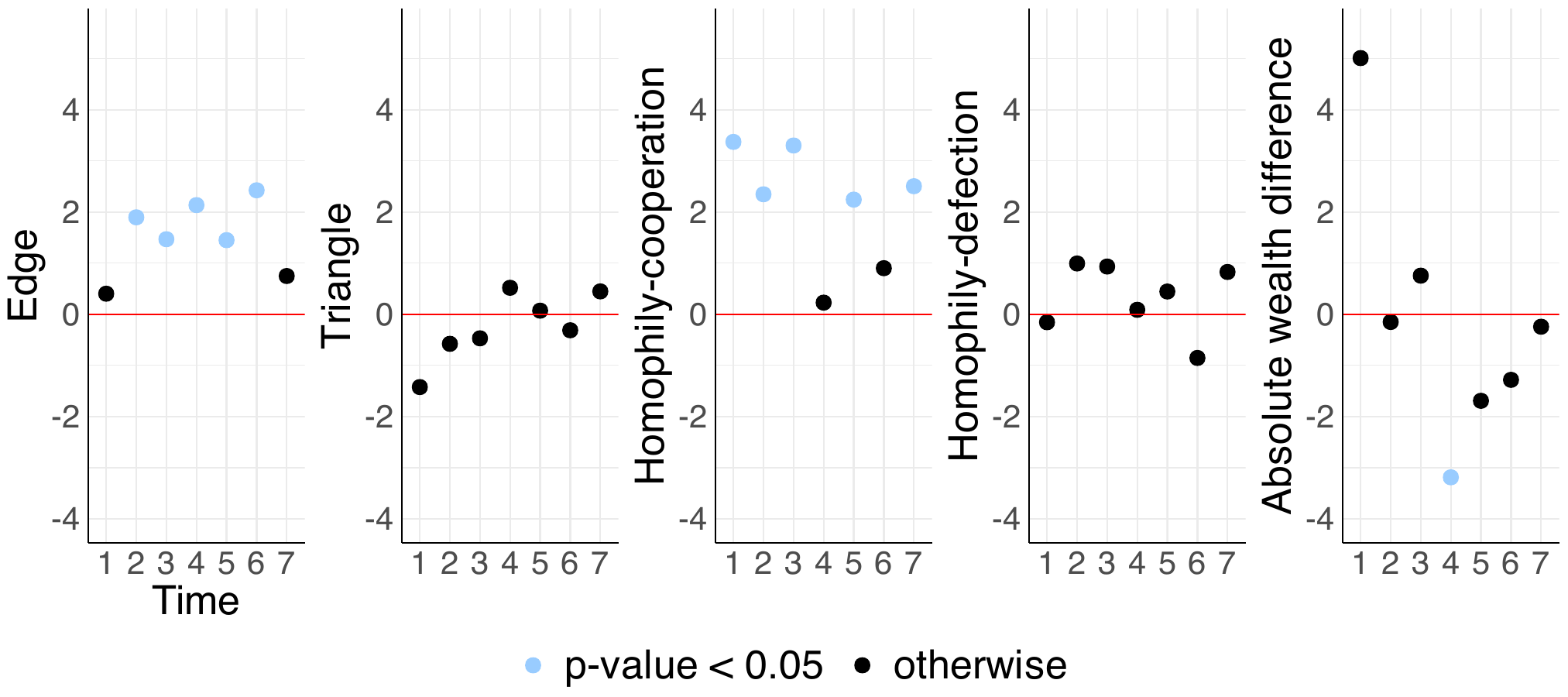}
\caption{MLE parameter estimates for the persistence model over time\label{fig6}}
\end{figure}

Finally, we conducted a robustness check to evaluate the homogeneity assumption of parameters across time. Fig. \ref{fig5} presents the MLE parameter estimates for the formation model over time, with colors indicating the significance of each parameter estimate (green: $p-$value $<$ 0.05, black: not significant). Consistent with the homogeneity model in Table \ref{tab1}, each parameter estimate demonstrates similar tendencies over time.
However, as there are no defections at time 6, the MLE of that parameter is negative infinity (and it is not plotted).
Fig. \ref{fig6} illustrates the MLEs for the parameters of the persistence model over time. As in Fig. \ref{fig5}, the colors represent the significance of each parameter estimate (blue: $p-$value $<$ 0.05; black: not significant). Again, the result aligns with the homogeneity model findings in Table 1, showing consistent tendencies for each parameter estimate across time.

\section{Discussion}\label{sec6}

In this paper, we proposed a statistical framework for analyzing networked public goods games using Separable Temporal Exponential-family Random Graph Models.  We demonstrated the application of this model by analysing an experiment of 20 games, highlighting the models ability to capture dependencies in temporal relational information.  The model provided novel insights into the formation and persistence of connections, uncovering mechanisms that drive network dynamics in social settings.  These insights, which could not have been achieved through nodal-level analysis alone, underscore the importance of examining dyad-level dependencies in understanding networked human behaviors.

In our proposed model, we employed direct numerical optimization of the likelihood function to achieve more reliable parameter estimates and likelihood ratios.  This approach is particularly advantageous for multiple small dynamic networks, as maximum likelihood estimation enables robust statistical inference and model comparison. By incorporating a number of independent networks, the model also maintains high statistical power.  However, the computational constraints of this method limits its applicability, making it suitable primarily for small-scale networks, particularly those with fewer than ten nodes.

Our proposed statistical framework supports several extensions.  For example, in evolutionary game theory, incorporating multiple treatment conditions can shed light on how experimental variables influence network dynamics.  As demonstrated
by \citet{ref18}, wealth visibility can affect cooperation rates in public goods games.  By including treatment conditions as covariates, the model can analyze the impact of different experimental setups on tie formation and persistence, enabling a deeper understanding of structural dependencies in social human behaviors.

In these experiments, the network size is controlled by the experimenter. However, it is a fundamental determinant of the social structure of the network, and hence the model terms and parameters. The models we have are conditional on the network size and direct comparison of the parameters for different network sizes is not directly possible \citep[See][]{ref14}.
A current limitation of the model is its inability directly compare experiments of different network sizes.  In social human behavior research, network sizes often fluctuate across networks, partly due to challenges in recruiting the constant number of participants for repeated experiments. Addressing this limitation, future research could focus on adapting the model to handle networks of varying sizes by utilizing methods such as those developed by \citet{ref14}. Incorporating these approaches would allow for more flexible model specifications and broaden the model’s applicability across diverse experimental settings.

Our proposed model holds potential for broad application across various domains, including networked human behaviors within families, workplaces, schools, and hospitals. In these settings, small groups of 6–10 individuals repeatedly interact over dynamic social networks. This model's versatility suggests its utility beyond the scope of networked public goods games, providing a pathway for advancements in understanding social interactions and behavioral dynamics across a variety of disciplines.

\section{Disclosure Statement}

The authors report there are no competing interests to declare.

\section{Data and Software Availability}
All analysis was done in the R environment (R version 4.3.2).
The code and available data to reconstruct the
analyses of this paper are available at
\url{https://github.com/Ankoudon/stergm-pgg}.

\section{Acknowledgements}

This article is based upon work supported by the National Science Foundation (NSF, 2230125, and CCF-2200197), Centers for Disease Control and National Institute of Child Health, Human Development (NICHD, R21HD063000, R21HD075714 and R24-HD041022), National Institutes of Health (NIH, K01AI166347), and the Japan Science and Technology Agency (JPMJPR21R8 and JPMJRS22K1). The content is solely the responsibility
of the authors and do not necessarily represent the official views of the National Institutes of Health or the National Science Foundation.

\section{Supplementary Materials}

\begin{description}

\item[Supplement] The supplement contains a descriptions of Exponential-family Random Graph Models, Temporal Exponential-family Random Graph Models, and some details of Separable Temporal Exponential-family Random Graph Models. (pdf) 

\end{description}

\section{Author Contributions}

HA and MSH designed the project. HA implemented the experiment and conducted the
data analysis. HA and MSH collaboratively wrote the paper. AN contributed laboratory resources. 

\bibliographystyle{chicago}

\bibliography{ms}

\end{document}



\newtheorem{assumption}{Assumption}
\newtheorem{definition}{Definition}
\newcommand{\R}{\mathbb{R}}
\newcommand{\N}{\mathbb{N}}
\newcommand{\E}{\mathbb{E}}
\newcommand{\V}{\mathbb{V}}
\newcommand{\bfR}{\mathbf{R}}
\newcommand{\bfX}{\mathbf{X}}
\newcommand{\bfW}{\mathbf{W}}
\newcommand{\bfD}{\mathbf{D}}
\newcommand{\INT}{\int_{-\infty}^{+\infty}}
\newcommand{\p}{\partial}
\newcommand{\ra}{\Rightarrow}
\newcommand{\dH}{d\mathscr{H}}
\newcommand{\ch}{\text{cosh}}
\newcommand{\sh}{\text{sinh}}
\newcommand{\ex}{\mathbb{E}\left[X\right]}
\newcommand{\ey}{\mathbb{E}\left[Y\right]}
\newcommand{\indep}{\perp \!\!\! \perp }

\setcounter{secnumdepth}{4}

\section{Exponential-family Random Graph Models}

This section gives a very brief introduction to Exponential-family Random Graph Models (ERGMs), to help
understand why they are appropriate for modeling the structure of a
complex social process that evolves over time.

ERGMs have a long history of
successfully representing dependencies in relational information \citep{ref7, ref10, ref23, ref21, ref1}.  Suppose that $n$ is the set of
actors/"nodes", let $\mathbf{Y}$ be an undirected random graph whose
realization is $\mathbf{y} \in \mathcal{Y}$, the set of possible networks of
interest on $n$.  With a d-vector $Z(\mathbf{y})$ of sufficient statistics and
parameter $\boldsymbol{\theta} \in \mathbb{R}^d$, an ERGM is expressed as

\begin{equation}
P(\mathbf{Y}=\mathbf{y}) = \exp\{\boldsymbol{\theta}{\cdot}Z(\mathbf{y}) - \psi(\boldsymbol{\theta})\} \quad \mathbf{y} \in \mathcal{Y},
\end{equation}
where
\begin{equation}
\exp\{\psi(\boldsymbol{\theta})\} = \sum_{\mathbf{x} \in \mathcal{Y}} \exp\{\boldsymbol{\theta}{\cdot}Z(\mathbf{x})\}
\end{equation}
is the normalizing constant. It is well known that the sum of $\exp\{\boldsymbol{\theta}{\cdot}Z(\mathbf{x})\}$ over the set of possible networks, $\mathcal{Y}$ often causes computational challenges; the number of possible networks on $n$ is $2^{n(n-1)/2}$, which is an astronomically large number for even moderate size $n$. Therefore, evaluating the log-likelihood, or even maximizing the log-likelihood, is computationally infeasible for large networks, resulting in the use of Markov Chain Monte Carlo (MCMC) methods \citep{ref6} to estimate the parameters of interest and conduct statistical inference. Notable here is the difficulty in reliably estimating the maximum likelihood estimators and its standard errors and implementing the likelihood ratio test. There are various reasons for this, such as the failure to generate well-mixed chains, the existence of MCMC errors, and uncertainty in approximation of the likelihood ratios \citep{ref10}.

It is a natural extension to apply the ERGM framework to representing dependencies in temporal relational information. Originating from \citet{ref20}, \citet{ref9} and \citet{ref8} defined the Temporal Exponential-family Random Graph Models (TERGMs) as ERGMs for the transition probability from time t to time t+1. More importantly, the introduction on the concept of separability of dynamic networks into formation and persistence networks has substantially improved interpretability and model specification through Separable TERGMs (STERGMs) \citep{ref13}. In these TERGM frameworks, parameter estimation and statistical inference are usually performed using MCMC methods as in cross-sectional ERGMs.

\section{Temporal Exponential-family Random Graph Model}\label{sec2}

\subsection{Model definition}\label{subsec2.1}

TERGMs are the natural extension of ERGMs. They were first introduced to model the network at time t conditional on the network at time t - 1 \citep{ref20, ref9}. Assuming that N is the set of nodes of interest, let $\mathbf{Y}^t$ be an undirected random graph at time t, whose realization is $\mathbf{y}^{t} \in \mathcal{Y}$, the set of possible networks of interest on N. With a d-vector $g(\mathbf{y}^{t}, \mathbf{y}^{t-1})$ of sufficient statistics for the network transition from $\mathbf{y}^{t-1}$ to $\mathbf{y}^{t}$ and parameter $\boldsymbol{\theta} \in \mathbb{R}^d$, the transition probability from time t - 1 to time t is defined as:
\begin{equation}
P(\mathbf{Y}^{t}=\mathbf{y}^{t}|\mathbf{Y}^{t-1}=\mathbf{y}^{t-1} ; \boldsymbol{\theta}) = \frac{\exp\{\boldsymbol{\theta}{\cdot}g(\boldsymbol{y}^{t}, \boldsymbol{y}^{t-1})\}}{c(\boldsymbol{\theta}, \boldsymbol{y}^{t-1})} \quad \boldsymbol{y}^{t}, \boldsymbol{y}^{t-1} \in \mathcal{Y},
\end{equation}
where
\begin{equation}
c(\boldsymbol{\theta}, \mathbf{y}^{t-1}) = \sum_{\mathbf{x}^t \in \mathcal{Y}} \exp\{\boldsymbol{\theta}{\cdot} g(\mathbf{x}^{t}, \mathbf{y}^{t-1})\}
\end{equation}
is the normalizing constant. An essential ingredient for model specification is the choice of g, which can be any valid network statistics evaluated at t that depends on t - 1. As a result, assuming constant parameters over time, a TERGM with T time points can be represented as:
\begin{equation}
\prod_{t=2}^{T}P(\mathbf{Y}^{t}=\mathbf{y}^{t}|\mathbf{Y}^{t-1}=\mathbf{y}^{t-1} ; \boldsymbol{\theta}).
\end{equation}

\subsection{Interpretation}\label{subsec2.2}

Similar to ERGMs, the parameters of TERGMs can be interpreted as the conditional odds. Given the property of conditional dyadic independence \citep{ref9}, the transition probability from time t - 1 to time t is re-expressed as:
\begin{equation}
P(\mathbf{Y}^{t}=\mathbf{y}^{t}|\mathbf{Y}^{t-1}=\mathbf{y}^{t-1} ; \boldsymbol{\theta}) = \prod_{i<j} P(\mathbf{Y}^{t}_{ij}=\mathbf{y}^{t}_{ij}|\mathbf{Y}^{t-1}=\mathbf{y}^{t-1} ; \boldsymbol{\theta}), 
\end{equation}
which means, in $\mathbf{Y}^t$, the realizations of tie states, $\mathbf{y}^{t}_{ij} $ are independent conditional on $\mathbf{Y}^{t-1}$, leading to the following model expression on the conditional odds:
\begin{equation}
\frac{P(\mathbf{Y}^{t}_{ij}=1|\mathbf{Y}^{t-1}=\mathbf{y}^{t-1}; \boldsymbol{\theta})}{P(\mathbf{Y}^{t}_{ij}=0|\mathbf{Y}^{t-1}=\mathbf{y}^{t-1}; \boldsymbol{\theta})} = \exp[\boldsymbol{\theta}{\cdot}\{g(\mathbf{y}_{+ij}^{t}, \mathbf{y}^{t-1}) - g(\mathbf{y}_{-ij}^{t}, \mathbf{y}^{t-1})\}] 
\end{equation}
here, $g(\mathbf{y}_{+ij}^{t}, \mathbf{y}^{t-1})$ is defined as a d-vector $g(\mathbf{y}^{t}, \mathbf{y}^{t-1})$ of sufficient statistics for the network transition from $\mathbf{y}^{t-1}$ to $\mathbf{y}^{t}$, where the edge $\mathbf{y}_{ij}^{t}$ is present, and $g(\mathbf{y}_{-ij}^{t}, \mathbf{y}^{t-1})$ is defined as a d-vector $g(\mathbf{y}^{t}, \mathbf{y}^{t-1})$ of sufficient statistics for the network transition from $\mathbf{y}^{t-1}$ to $\mathbf{y}^{t}$, where the edge $\mathbf{y}_{ij}^{t}$ is absent. This interpretation is crucial for understanding the dependencies in the network transition and the role of the parameters in the model.

However, it is important to note that in this model, both the interpretation of the parameters and the model specification may present challenges \citep{ref13}. For example, when interpreting dyadic homophily statistics given certain nodal-level groupings, the statistics can be defined as:
\begin{equation}
g(\mathbf{z}, \mathbf{y}^{t}, \mathbf{y}^{t-1}) = \sum_{i<j} \mathbf{z}_{ij}\mathbf{y}_{ij}^{t}, 
\end{equation}
where, the nodal-level groupings are defined as:
\begin{equation}
\mathbf{z}_{ij} = 
\begin{cases} 
    1 & \text{the node i and j are in the same group} \\
    0 & \text{otherwise} 
\end{cases}
\end{equation}
A higher value of the corresponding parameter indicates that more ties are likely to be present between nodes within the same group in the realizations, $\mathbf{y}^{t} \in \mathcal{Y}$. Conversely, a lower value of the parameter suggests that fewer ties are likely to be present between nodes within the same group in the realizations, $\mathbf{y}^{t} \in \mathcal{Y}$. Still, it is important to recognize that these dynamic processes occur through the simultaneous formation and dissolution of ties: with a higher parameter value, the dyads might be toggled 'on' more if they were previously empty (indicating more formation) and be toggled 'off' less if they were already present (indicating less dissolution), and vice versa. In this respect, the model is limited in that it cannot distinguish between the formation and dissolution of ties, which poses a challenge in interpreting the parameters.

In addition, the primary challenge of this modeling framework becomes evident through insights into the further inspection above, which directly influence the model specification. Incorporating the dyadic homophily statistics and its parameter, $\theta_0$, the the transition probability of node i and j from time t - 1 to time t can be expressed as follows:
\begin{align}
&\frac{P(\mathbf{Y}^{t}_{ij}=1|\mathbf{Y}^{t-1}=\mathbf{y}^{t-1}, \mathbf{Y}^{t-1}_{ij}=0, \mathbf{z}_{ij} = 1; \theta_{0})}{P(\mathbf{Y}^{t}_{ij}=0|\mathbf{Y}^{t-1}=\mathbf{y}^{t-1}, \mathbf{Y}^{t-1}_{ij}=0, \mathbf{z}_{ij} = 1; \theta_{0})} \\
&= \frac{P(\mathbf{Y}^{t}_{ij}=1|\mathbf{Y}^{t-1}=\mathbf{y}^{t-1}, \mathbf{Y}^{t-1}_{ij}=1, \mathbf{z}_{ij} = 1; \theta_{0})}{P(\mathbf{Y}^{t}_{ij}=0|\mathbf{Y}^{t-1}=\mathbf{y}^{t-1}, \mathbf{Y}^{t-1}_{ij}=1, \mathbf{z}_{ij} = 1; \theta_{0})} \\
&= \exp(\theta_0) \nonumber \\
\Rightarrow
&\begin{cases} 
P(\mathbf{Y}^{t}_{ij}=1|\mathbf{Y}^{t-1}=\mathbf{y}^{t-1}, \mathbf{Y}^{t-1}_{i,j}=0, \mathbf{z}_{ij} = 1; \theta_{0}) = \frac{\exp(\theta_0)}{1+\exp(\theta_0)} \\
P(\mathbf{Y}^{t}_{ij}=0|\mathbf{Y}^{t-1}=\mathbf{y}^{t-1}, \mathbf{Y}^{t-1}_{i,j}=1, \mathbf{z}_{ij} = 1; \theta_{0})= 1-\frac{\exp(\theta_0)}{1+\exp(\theta_0)}.
\end{cases}
\end{align}
Thus, with a higher parameter value, the dyads are more likely to be toggled 'on' if they were previously empty (indicating more formation) and less likely to be toggled 'off' if they were already present (indicating less dissolution), and with a lower parameter value, the dyads are less likely to be toggled ‘on’ if they were previously empty (indicating less formation) and more likely to be toggled ‘off’ if they were already present (indicating more dissolution). This is a significant limitation of the model, as it can only capture the overall dynamics of the network transitions in specific ways, rather than distinguishing between the formation and dissolution of dynamics.

\section{Separable Temporal Exponential-family Random Graph Model}\label{sec3}
\subsection{Model definition}\label{subsec3.1}

Separable Temporal Exponential-family Random Graph Models (STERGMs) were introduced by \citet{ref13} as a subset of TERGMs for better interpretability and model specification. The main concept is to "separate" the dynamic network into distinct formation and persistence processes. 

Consider the network transition from time t -1 to time t, defining the network $\mathbf{Y}^{t-1}$ at time t -1, the network $\mathbf{Y}^{t}$ at time t, the formation network $\mathbf{Y}^+$; the initial network $\mathbf{Y}^{t-1}$ with the addition of ties at time t, and the persistence network $\mathbf{Y}^-$; the initial network $\mathbf{Y}^{t-1}$ with the removal of ties at time t. Via a set operation, the realized formation and persistence networks are derived as:
\begin{equation}
\begin{aligned}
  &\mathbf{y}^+ = \mathbf{y}^{t-1} \cup \mathbf{y}^t  \\
  &\mathbf{y}^- = \mathbf{y}^{t-1} \cap \mathbf{y}^t
\end{aligned}
\end{equation}
In this operation, $\mathbf{y}^+ = \mathbf{y}^{t-1} \cup \mathbf{y}^t$ represents the set of ties that appear in either the network at time t - 1 or the network at time t. Conversely, $\mathbf{y}^- = \mathbf{y}^{t-1} \cap \mathbf{y}^t$ represents the set of ties that exist in both the network at time t - 1 and the network at time t. A key goal of STERGMs is to reconstruct $\mathbf{y}^t$ from $\mathbf{y}^{t-1}$, $\mathbf{y}^+$, and $\mathbf{y}^-$, or to separate $\mathbf{y}^t$ into $\mathbf{y}^+$ and $\mathbf{y}^-$, given $\mathbf{y}^{t-1}$. This reconstruction is achieved with the following set operation:
\begin{equation}
\mathbf{y}^t = \mathbf{y}^+ \backslash (\mathbf{y}^{t-1} \backslash \mathbf{y}^-) = \mathbf{y}^- \cup (\mathbf{y}^+ \backslash \mathbf{y}^{t-1}), 
\end{equation}
where, $\mathbf{y}^+ \backslash \mathbf{y}^{t-1}$ contains ties $\{i, j\}$  that are present in $\mathbf{y}^+$ but not in $\mathbf{y}^{t-1}$. Thus, $\mathbf{y}^t$ can be expressed as the union of $\mathbf{y}^-$ and $\mathbf{y}^+ \backslash \mathbf{y}^{t-1}$. This approach allows us to separate the processes of ties into the formation and the persistence as the network evolves over time. As a result, if $\mathbf{Y}^+$ is independent of $\mathbf{Y}^-$ conditional on $\mathbf{Y}^{t-1}$, the transition probability from time t - 1 to time t is separable as follows:
\begin{align}
&P(\mathbf{Y}^t = \mathbf{y}^t | \mathbf{Y}^{t-1} = \mathbf{y}^{t-1}; \boldsymbol{\theta}) = P(\mathbf{Y}^+ = \mathbf{y}^+, \mathbf{Y}^- = \mathbf{y}^- | \mathbf{Y}^{t-1} = \mathbf{y}^{t-1}; \boldsymbol{\theta}^+, \boldsymbol{\theta}^-) \nonumber \\
&=P(\mathbf{Y}^+ = \mathbf{y}^+ | \mathbf{Y}^{t-1} = \mathbf{y}^{t-1}; \boldsymbol{\theta}^+) \times P(\mathbf{Y}^- = \mathbf{y}^- | \mathbf{Y}^{t-1} = \mathbf{y}^{t-1}; \boldsymbol{\theta}^-).
\end{align}

Specifically, we respectively model the formation and the persistence models. Given $\boldsymbol{y}^{t-1} \in \mathcal{Y}$, the realizations of $\boldsymbol{Y}^+$ can be expressed as $\boldsymbol{y}^+ \in \mathcal{Y}^+(\boldsymbol{y}^{t-1}) \subseteq\{\forall \boldsymbol{y} : \boldsymbol{y} \supseteq \boldsymbol{y}^{t-1}\}$ and the realizations of $\boldsymbol{Y}^-$ is expressed as $\boldsymbol{y}^- \in \mathcal{Y}^-(\boldsymbol{y}^{t-1}) \subseteq\{\forall \boldsymbol{y} : \boldsymbol{y} \subseteq \boldsymbol{y}^{t-1}\}$. With a d-vector $g^+(\boldsymbol{y}^{+}, \boldsymbol{y}^{t-1})$ of sufficient statistics for the formation network $\boldsymbol{y}^+$ from $\boldsymbol{y}^{t-1}$ and parameter $\boldsymbol{\theta}^+ \in \mathbb{R}^d$ and a d-vector $g^-(\boldsymbol{y}^{-}, \boldsymbol{y}^{t-1})$ of sufficient statistics for the persistence network $\boldsymbol{y}^-$ from $\boldsymbol{y}^{t-1}$ and parameter $\boldsymbol{\theta}^- \in \mathbb{R}^d$, the formation and persistence models are elaborated as:
\begin{align}
P(\boldsymbol{Y}^+ = \boldsymbol{y}^+ | \boldsymbol{Y}^{t-1} = \boldsymbol{y}^{t-1}; \boldsymbol{\theta}^+) = \frac{\exp(\boldsymbol{\theta}^{+} {\cdot}g^+(\boldsymbol{y}^{+}, \boldsymbol{y}^{t-1}))}{c^+(\boldsymbol{\theta}^+, \boldsymbol{y}^{t-1})} \quad \boldsymbol{y}^+ \in \mathcal{Y}^+(\boldsymbol{y}^{t-1}), \\
P(\boldsymbol{Y}^- = \boldsymbol{y}^- | \boldsymbol{Y}^{t-1} = \boldsymbol{y}^{t-1}; \boldsymbol{\theta}^-) = \frac{\exp(\boldsymbol{\theta}^{-} {\cdot}g^-(\boldsymbol{y}^{-}, \boldsymbol{y}^{t-1}))}{c^-(\boldsymbol{\theta}^-, \boldsymbol{y}^{t-1})} \quad \boldsymbol{y}^- \in \mathcal{Y}^-(\boldsymbol{y}^{t-1}),
\end{align}
where 
\begin{align}
c^+(\boldsymbol{\theta}^+, \boldsymbol{y}^{t-1}) &= \sum_{\boldsymbol{x}^+ \in \mathcal{Y}^+(\boldsymbol{y}^{t-1})} \exp\{\boldsymbol{\theta}^{+} {\cdot}g^+(\boldsymbol{x}^{+}, \boldsymbol{y}^{t-1})\}, \\
c^-(\boldsymbol{\theta}^-, \boldsymbol{y}^{t-1}) &= \sum_{\boldsymbol{x}^- \in \mathcal{Y}^-(\boldsymbol{y}^{t-1})} \exp\{\boldsymbol{\theta}^{-} {\cdot}g^-(\boldsymbol{x}^{-}, \boldsymbol{y}^{t-1})\},
\end{align}
are the normalizing constants. In this framework, the sufficient statistics for the formation and persistence networks can vary, allowing for a more flexible model specification \citep{ref13}. In practice, this property is considered to be useful \citep{ref11, ref12}. For instance, in an extreme case, the formation network model might include statistics that capture homophily ties, while the persistence network does not. Although STERGMs sacrifice the ability to model interactions between the formation and persistence networks, it offers significant improvements in model specification and interpretability. Finally, we demonstrate that STERGMs form a subclass of TERGMs as follows:
\begin{align}
&P(\boldsymbol{Y}^+ = \boldsymbol{y}^+ | \boldsymbol{Y}^{t-1} = \boldsymbol{y}^{t-1}; \boldsymbol{\theta}^+) \times P(\boldsymbol{Y}^- = \boldsymbol{y}^- | \boldsymbol{Y}^{t-1} = \boldsymbol{y}^{t-1}; \boldsymbol{\theta}^-) \nonumber \\
&= \frac{\exp(\boldsymbol{\theta}^{+}{\cdot} g^+(\boldsymbol{y}^{+}, \boldsymbol{y}^{t-1}))}{c^+(\boldsymbol{\theta}^+, \boldsymbol{y}^{t-1})} \cdot \frac{\exp(\boldsymbol{\theta}^{-} {\cdot}g^-(\boldsymbol{y}^{-}, \boldsymbol{y}^{t-1}))}{c^-(\boldsymbol{\theta}^-, \boldsymbol{y}^{t-1})} \nonumber \\
&= \frac{\exp(\boldsymbol{\theta}^{+} {\cdot}g^+(\boldsymbol{y}^{+}, \boldsymbol{y}^{t-1}) + \boldsymbol{\theta}^{-} {\cdot}g^-(\boldsymbol{y}^{-}, \boldsymbol{y}^{t-1}))}{c^+(\boldsymbol{\theta}^+, \boldsymbol{y}^{t-1}) \cdot c^-(\boldsymbol{\theta}^-, \boldsymbol{y}^{t-1})} \nonumber \\
&= \frac{\exp\{(\boldsymbol{\theta}^{+}, \boldsymbol{\theta}^{-}) {\cdot}(g^+(\boldsymbol{y}^{+}, \boldsymbol{y}^{t-1}), g^-(\boldsymbol{y}^{-}, \boldsymbol{y}^{t-1}))\}}{\sum_{\boldsymbol{x}^+ \in \mathcal{Y}^+(\boldsymbol{y}^{t-1}), \boldsymbol{x}^- \in \mathcal{Y}^-(\boldsymbol{y}^{t-1})} \exp\{(\boldsymbol{\theta}^{+}, \boldsymbol{\theta}^{-}) {\cdot}(g^+(\boldsymbol{x}^{+}, \boldsymbol{y}^{t-1}), g^-(\boldsymbol{x}^{-}, \boldsymbol{y}^{t-1}))\}} \nonumber \\
&= \frac{\exp\{(\boldsymbol{\theta}^{+}, \boldsymbol{\theta}^{-}) (g^+(\boldsymbol{y}^{t-1} \cup \boldsymbol{y}^t, \boldsymbol{y}^{t-1}), g^-(\boldsymbol{y}^{t-1} \cap \boldsymbol{y}^t, \boldsymbol{y}^{t-1}))\}}{\sum_{\boldsymbol{w}^t \in \mathcal{Y}} \exp\{(\boldsymbol{\theta}^{+}, \boldsymbol{\theta}^{-}) {\cdot}(g^+(\boldsymbol{y}^{t-1} \cup \boldsymbol{w}^t, \boldsymbol{y}^{t-1}), g^-(\boldsymbol{y}^{t-1} \cap \boldsymbol{w}^t, \boldsymbol{y}^{t-1}))\}} \nonumber \\
&= \frac{\exp\{\boldsymbol{\theta}^{*} {\cdot}g^*(\boldsymbol{y}^t, \boldsymbol{y}^{t-1})\}}{\sum_{\boldsymbol{w}^t \in \mathcal{Y}} \exp\{\boldsymbol{\theta}^{*} {\cdot}g^*(\boldsymbol{w}^t, \boldsymbol{y}^{t-1})\}}, 
\end{align}
where
\begin{equation}
\begin{cases}
\boldsymbol{\theta}^{*} = (\boldsymbol{\theta}^{+}, \boldsymbol{\theta}^{-}), \\
g^*(\boldsymbol{y}^t, \boldsymbol{y}^{t-1}) = (g^+(\boldsymbol{y}^{t-1} \cup \boldsymbol{y}^t, \boldsymbol{y}^{t-1}), g^-(\boldsymbol{y}^{t-1} \cap \boldsymbol{y}^t, \boldsymbol{y}^{t-1})).
\end{cases}
\end{equation}
The final form is identical to that of a TERGM, underscoring that STERGMs represent a specialized case within the broader TERGM framework. 

\subsection{Interpretation}\label{subsec3}

The parameters of STERGMs can be interpreted as conditional odds, as for TERGMs. Given the property of conditional dyadic independence (Hanneke and Xing, 2006), the formation and persistence models can be re-expressed as:
\begin{align}
&P(\boldsymbol{Y}^{+}=\boldsymbol{y}^{+}|\boldsymbol{Y}^{t-1}=\boldsymbol{y}^{t-1} ; \boldsymbol{\theta}^+) = \prod_{i<j} P(\boldsymbol{Y}^{+}_{ij}=\boldsymbol{y}^{+}_{ij}|\boldsymbol{Y}^{t-1}=\boldsymbol{y}^{t-1} ; \boldsymbol{\theta}^+), \\
&P(\boldsymbol{Y}^{-}=\boldsymbol{y}^{-}|\boldsymbol{Y}^{t-1}=\boldsymbol{y}^{t-1} ; \boldsymbol{\theta}^-) = \prod_{i<j} P(\boldsymbol{Y}^{-}_{ij}=\boldsymbol{y}^{-}_{ij}|\boldsymbol{Y}^{t-1}=\boldsymbol{y}^{t-1} ; \boldsymbol{\theta}^-),
\end{align}
which means, in $\mathbf{Y}^+$, the realizations of tie states, $\mathbf{y}^{+}_{ij} $ are independent conditional on $\mathbf{Y}^{t-1}$, and in $\mathbf{Y}^-$, the realizations of tie states, $\mathbf{y}^{-}_{ij} $ are independent conditional on $\mathbf{Y}^{t-1}$, leading to the following model expression on the conditional odds:
\begin{align}
&\frac{P(\boldsymbol{Y}^{+}_{ij}=1|\boldsymbol{Y}^{t-1}=\boldsymbol{y}^{t-1}; \boldsymbol{\theta}^+)}{P(\boldsymbol{Y}^{+}_{ij}=0|\boldsymbol{Y}^{t-1}=\boldsymbol{y}^{t-1}; \boldsymbol{\theta}^+)} = \exp[\boldsymbol{\theta}^{+}{\cdot}\{g^+(\boldsymbol{y}_{+ij}^{+}, \boldsymbol{y}^{t-1}) - g^+(\boldsymbol{y}_{-ij}^{+}, \boldsymbol{y}^{t-1})\}], \\
&\frac{P(\boldsymbol{Y}^{-}_{ij}=1|\boldsymbol{Y}^{t-1}=\boldsymbol{y}^{t-1}; \boldsymbol{\theta}^-)}{P(\boldsymbol{Y}^{-}_{ij}=0|\boldsymbol{Y}^{t-1}=\boldsymbol{y}^{t-1}; \boldsymbol{\theta}^-)} = \exp[\boldsymbol{\theta}^{-}\{g^-(\boldsymbol{y}_{+ij}^{-}, \boldsymbol{y}^{t-1}) - g^-(\boldsymbol{y}_{-ij}^{-}, \boldsymbol{y}^{t-1})\}],
\end{align}
here, $g(\boldsymbol{y}_{+ij}^{+}, \boldsymbol{y}^{t-1})$ is defined as a d-vector $g(\boldsymbol{y}^{+}, \boldsymbol{y}^{t-1})$ of sufficient statistics for the network transition from $\boldsymbol{y}^{t-1}$ to $\boldsymbol{y}^{+}$, where the edge $\boldsymbol{y}_{ij}^{+}$ is present, and $g(\boldsymbol{y}_{-ij}^{+}, \boldsymbol{y}^{t-1})$ is defined as a d-vector $g(\boldsymbol{y}^{+}, \boldsymbol{y}^{t-1})$ of sufficient statistics for the network transition from $\boldsymbol{y}^{t-1}$ to $\boldsymbol{y}^{+}$, where the edge $\boldsymbol{y}_{ij}^{+}$ is absent. In the same manner, $g(\boldsymbol{y}_{+ij}^{-}, \boldsymbol{y}^{t-1})$ and $g(\boldsymbol{y}_{-ij}^{-}, \boldsymbol{y}^{t-1})$ can be also interpreted. 

For the formation model, a positive $\boldsymbol{\theta}^+_{k}$ indicates that one unit increase in $g^+(\boldsymbol{y}_{+ij}^{+}, \boldsymbol{y}^{t-1}) - g^+(\boldsymbol{y}_{-ij}^{+}, \boldsymbol{y}^{t-1})$ leads to a higher conditional log-odds of the edge $\boldsymbol{y}_{ij}^{+}$ being present, given $\boldsymbol{y}^{t-1}$. Conversely, a negative $\boldsymbol{\theta}^+_{k}$ indicates that one unit increase in $g^+(\boldsymbol{y}_{+ij}^{+}, \boldsymbol{y}^{t-1}) - g^+(\boldsymbol{y}_{-ij}^{+}, \boldsymbol{y}^{t-1})$ leads to a lower conditional log-odds of the edge $\boldsymbol{y}_{ij}^{+}$ being present, given $\boldsymbol{y}^{t-1}$.

For the persistence model, a positive $\boldsymbol{\theta}^-_{k}$ indicates that one unit increase in $g^-(\boldsymbol{y}_{+ij}^{-}, \boldsymbol{y}^{t-1}) - g^-(\boldsymbol{y}_{-ij}^{-}, \boldsymbol{y}^{t-1})$ leads to a higher conditional log-odds of the edge $\boldsymbol{y}_{ij}^{-}$ being present, given $\boldsymbol{y}^{t-1}$. Conversely, a negative $\boldsymbol{\theta}^-_{k}$ indicates that one unit increase in $g^-(\boldsymbol{y}_{+ij}^{-}, \boldsymbol{y}^{t-1}) - g^-(\boldsymbol{y}_{-ij}^{-}, \boldsymbol{y}^{t-1})$ leads to a lower conditional log-odds of the edge $\boldsymbol{y}_{ij}^{-}$ being present, given $\boldsymbol{y}^{t-1}$. 

\section{Author Contributions}

HA and MSH designed the project. HA implemented the experiment and conducted the
data analysis. HA and MSH collaboratively wrote the paper. AN contributed laboratory resources. 

\bibliographystyle{chicago}
\bibliography{supplement}